\documentclass{PoS}

\title{Status of ground-based gamma-ray astronomy}

\ShortTitle{Ground-based gamma-ray astronomy}

\author{\speaker{Marianne Lemoine-Goumard}\\%\thanks{A footnote may follow.}\\
        CENBG, CNRS-IN2P3, Universit\'e de Bordeaux\\
        19 Chemin du Solarium, CS 10120, F-33175 GRADIGNAN Cedex \\
        E-mail: \email{lemoine@cenbg.in2p3.fr}}

\abstract{
This article is the write-up of a rapporteur talk given at the 34th ICRC in The Hague, Netherlands. It attempts to review the results and developments presented at the conference and associated to the vibrant field of ground-based gamma-ray astronomy.  In total, it aims to give an overview of the 19 gamma-ray sessions, 84 talks and 176 posters presented at the 34th ICRC on this topic. New technical advances and projects will be described with an emphasis given on the cosmic-ray related studies of the Universe.}

\FullConference{The 34th International Cosmic Ray Conference,\\
		30 July- 6 August, 2015\\
		The Hague, The Netherlands}

\begin{document}
Very high-energy (VHE) electromagnetic radiation reaches Earth from a large part of the Cosmos, carrying crucial and unique
information about the most energetic phenomena in the Universe. Yet, it has only been in the last 25 years that we have had instruments
to "observe" this radiation. The situation changed with the development of imaging air Cherenkov telescopes (IACTs) and particle detectors, which have now matured to open a new window for exploration of the high-energy Universe. These 2 different strategies to detect gamma rays are very complementary: IACTs have a small field of view, a small duty cycle but very good angular and energy resolution, while particle detectors have a much larger field of view and duty cycle but poor angular resolution in comparison to the IACTs. These two types of detectors are also extremely complementary with (and benefit from) gamma-ray satellites such as \emph{Fermi}-LAT which is described in an accompanying article giving a
status report on space-based gamma-ray astronomy~\cite{rolf}.

\section{Status of the different experiments}
\label{status}
Currently, six main experiments are operational : the IACTs VERITAS, MAGIC, FACT and H.E.S.S. and the particle detectors HAWC, and Tibet AS$\gamma$. On the same site in Tibet, data-taking with ARGO-YBJ concluded after 5 years of observations in February 2013 but some summary results were presented at this ICRC.

\begin{figure}[b]
 \centering
     \includegraphics[height=0.3\textheight]{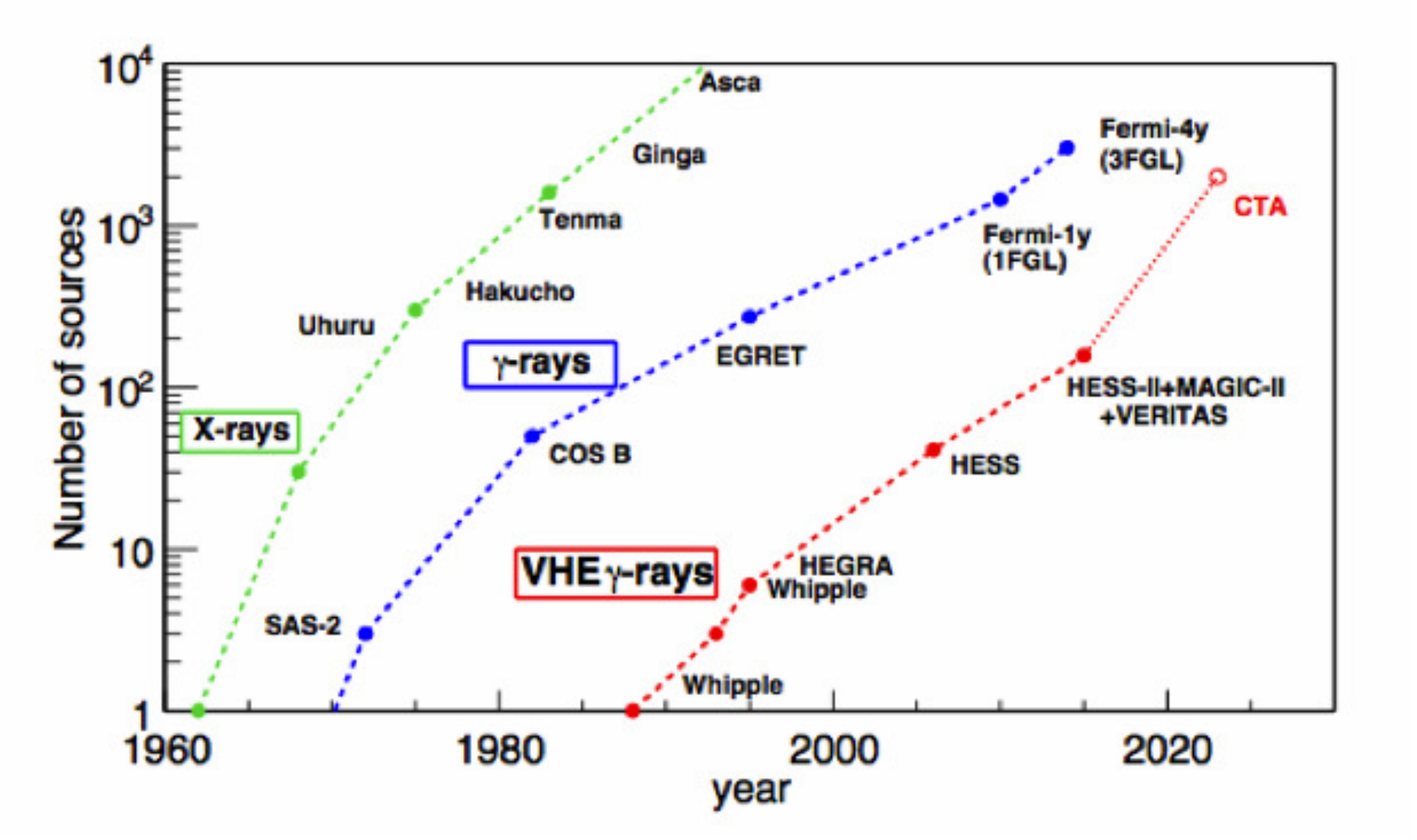}
  \caption{Kifune plot (named after T. Kifune, who first showed a similar plot at the 1995 ICRC in Rome), showing
the number of sources detected over time for various wavebands~\cite{denauroi}.
}
  \label{fig:tevsrcs}
\end{figure}

\subsection{Imaging Atmospheric Cherenkov Telescopes}
VERITAS, at the Fred Lawrence Whipple Observatory in southern Arizona has 4 telescopes of 12m diameter and cameras, covering a FoV of 3.5$^{\circ}$~\cite{staszak}. Since inauguration, the array has undergone two major hardware upgrades. The first, in 2009, relocated one of the telescopes to better symmetrize the array, while the second, over 2011-2012, replaced the L2 trigger system and instrumented each of the cameras with new, high efficiency PMTs. Additionally, VERITAS collect data under bright moonlight conditions using two non-standard techniques, one with reduced high voltage PMT settings and the other with UV filters placed over the PMTs. The utility of adding observation time under these conditions was demonstrated by the detection of a flare state in a BL Lac object, 1ES 1727+502, under bright moonlight conditions~\cite{cerutti}.\\

MAGIC is a system of two 17m diameter IACTs located at 2200m at La Palma on the Canaries Island. Between summer 2011 and 2012 the telescopes went through a major upgrade, carried out in two stages~\cite{sitarek}. In the first part of the upgrade, in summer 2011, the readout systems of both telescopes were upgraded. In summer 2012, the second stage of the upgrade followed with an exchange of the camera of the MAGIC-I telescope to a uniformly pixellized one. Finally, during Winter 2013-2014, after several years of development, a new system (sum-trigger) was implemented for stereoscopic observations. The energy threshold with this sum-trigger is 35 GeV. \\

Located on the same site as MAGIC, the FACT telescope consists of a 1440-pixel G-APD camera at the focus of one of the original HEGRA telescopes~\cite{biland}. Since Summer 2012, FACT is operated remotely (without the need of a data-taking crew on site). Even more important, during more than three years of operation of FACT, G-APDs have proven to be very reliable. The results presented at this ICRC, in particular the spectrum of the Crab nebula in excellent agreement with other IACTs~\cite{temme}, are a clear proof of the reliability, stability and performance of this system, showing that they are an alternative and excellent solution for the future.\\

H.E.S.S. is originally an array of 4 12m diameter telescopes. Since 2012, a fifth telescope of 28m located at the center of the array is operational. This allows to lower the energy threshold down to 30 GeV at zenith. Since the camera electronics of CT1-4 are much older than the one of CT5, an upgrade is being carried out since summer 2015 with a full completion planned for 2016. The goals of this upgrade are threefold: reducing the dead time of the cameras, improving the overall performance of the array and reducing the system failure rate related to aging~\cite{giavitto}.\\
 
\subsection{Particle detectors}
The High Altitude Water Cherenkov (HAWC) Gamma-Ray Observatory, located at Sierra Negra, Mexico at 4100 m a.s.l., is sensitive to gamma rays in the energy range of 100 GeV to 100 TeV~\cite{pretz}. HAWC consists of an array of 300 water Cherenkov detectors (WCDs) and a predicted point source sensitivity of 10 times that of its predecessor Milagro. The full HAWC array was completed in March 2015, but science operations already started in August 2013 with a partially-built array. As said above, one of the real advantage of particle detectors such as HAWC is their large field of view and high duty cycle.\\

Two other large particle detector arrays are located at very high altitude, 4300 m a.s.l., in Yangbajing (Tibet, China): ARGO-YBJ and Tibet AS$\gamma$. ARGO-YBJ consists of a single layer of resistive plate chambers completely covering an area of 100 $\times$ 110 m. It has been operated stably for 5.3 years, with an average duty cycle of 86\% and for a total effective time of 1670.45 days~\cite{argo_diffuse}. The Tibet AS$\gamma$ air shower array, also at Yangbajing, is still operationnal~\cite{tibet_crab}. It consists of 789 closely-scintillator detectors covering an area of 36900 m2. It is in operation since 1999 but a recent upgrade with a muon detector array will allow to reduce the background by selecting gamma-like events. Simulations demonstrate that using the full-scale MD array will enable to reject background cosmic-ray events by $\sim$99.99\% at 100 TeV, and that the sensitivity of the AS array to 10--1000 TeV gamma-ray sources will be improved by more than an order of magnitude. Currently, five modules among the 12 are operational and data-taking with this array in started 2014.\\

\begin{figure}[t]
 \centering
     \includegraphics[height=0.27\textheight]{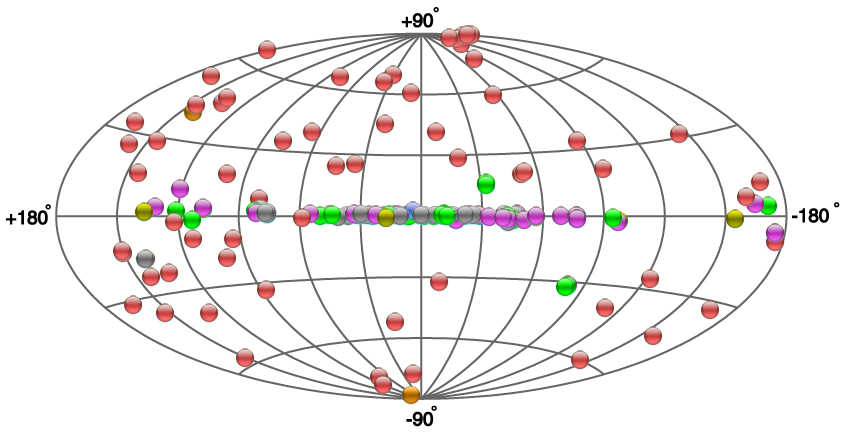}
     \includegraphics[height=0.25\textheight]{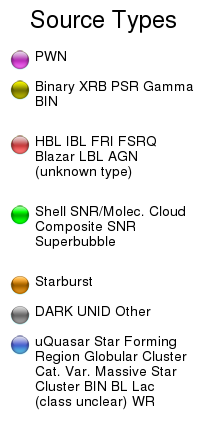}
  \caption{The TeV source catalog in Galactic coordinates as of summer 2015, courtesy of TeVCat.
}
  \label{fig:tevcat}
\end{figure}

It is clear from this description that all experiments have carried out, or are still in the process of doing an upgrade to improve their sensitivity or even lower the energy threshold. These improved performances is one of the key that allows the exponential rise of the number of sources since 25 years now as can be seen in Figure~\ref{fig:tevsrcs}. As of today, 162 sources are reported in the TeV catalog : 65 are extragalactic, 70 are Galactic, 27 are unidentified, many of them being also of Galactic origin as can be seen from the distribution of the grey sources in Figure \ref{fig:tevcat} which are almost all located in the Galactic plane. Even more importantly, the field is moving beyond quantity towards a qualitatively better understanding about the high-energy processes at work in these sources. In the following, I will try to review the main highlights discussed at this ICRC. Unavoidably this selection is subjective, and incomplete.

\section{The Galactic sky}
A lot can happen in our violent neighborhood. Massive stars will end their lives in a terrible explosion called supernova which can lead to the formation of a pulsar or a black hole and a shock that propagates in the interstellar medium called a supernova remnant. All these different sources can accelerate particles to very high energy. In turn, these particles will emit gamma rays that can be observed by ground-based experiments.

\subsection{The Galactic Plane survey}

\begin{figure}[t]
 \centering
     \includegraphics[height=0.35\textheight]{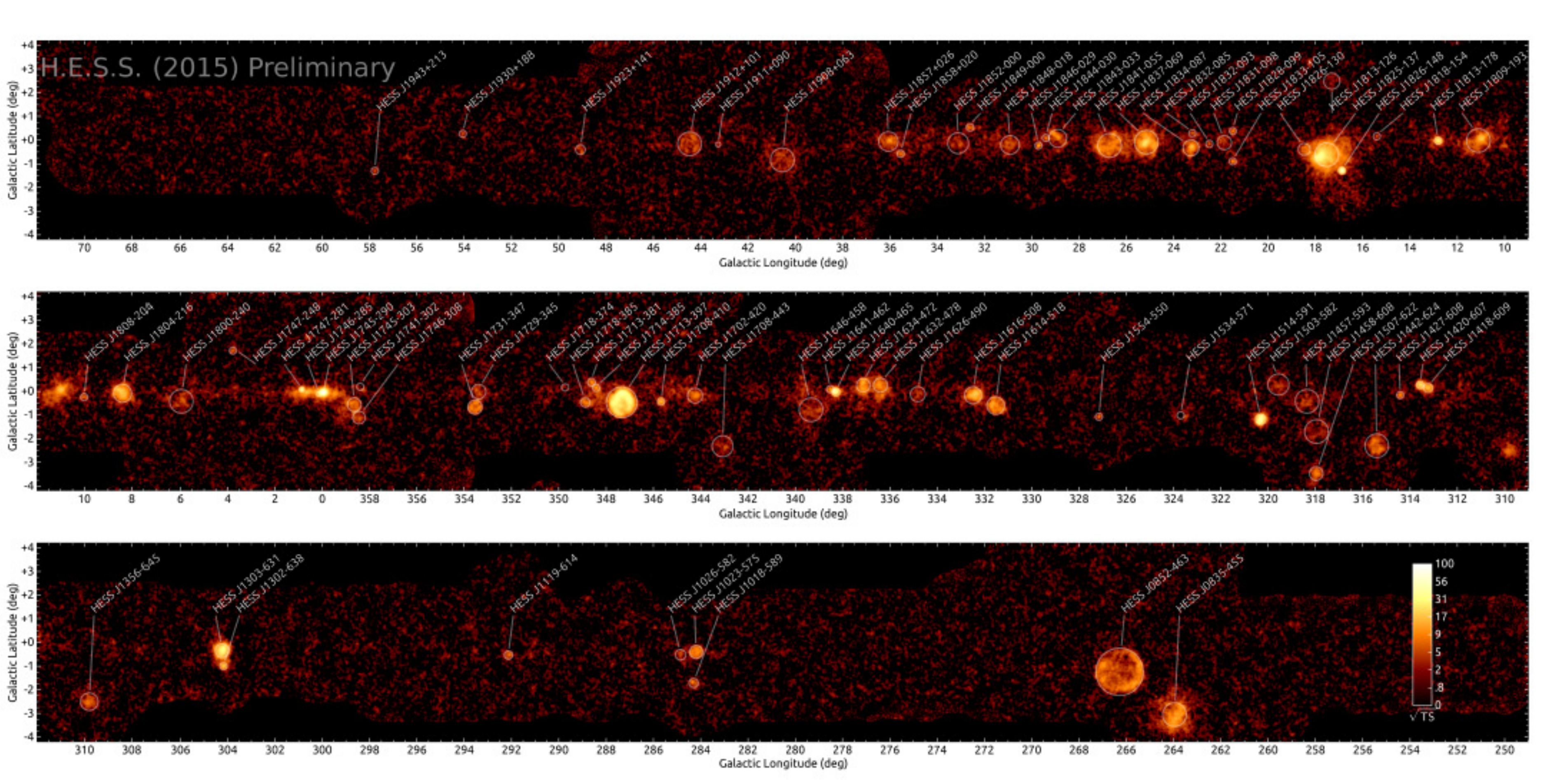}
  \caption{TS map for the H.E.S.S. Galactic Plane Survey~\cite{hgps}. Identifiers for sources that have been described in
publications or announced at conferences are included.
}
  \label{fig:hgps}
\end{figure}

The Galactic plane survey is certainly the investigation that provides the best general view of our Galaxy. Thanks to its location in the southern hemisphere, HESS has a privileged access to a very large part of the Galaxy. Roughly 2800 hours of high-quality observations of the Galactic disk are available in the Galactic longitude range 250 to 65 degrees and Galactic latitude range b $<$ 3.5 degrees. This is the first high-resolution ($\sim$0.1 deg) and sensitive ($\sim$2\% Crab nebula point-source sensitivity) survey of the Milky Way in TeV gamma-rays. The HESS Galactic Plane Survey (HGPS) has revealed a diverse population of cosmic accelerators in the Galaxy, from which 77 very-high-energy (E > 0.1 TeV) $\gamma$-ray sources have been compiled thanks to a new pipeline developed in this context for both source detection and characterization (morphology and spectrum)~\cite{hgps}. Out of the 77 sources, 16 are new sources that were previously unknown or unpublished. The distribution of the TeV sources along the Galactic Plane is presented in Figure~\ref{fig:hgps}.

\begin{figure}[h]
 \centering
     \includegraphics[height=0.2\textheight]{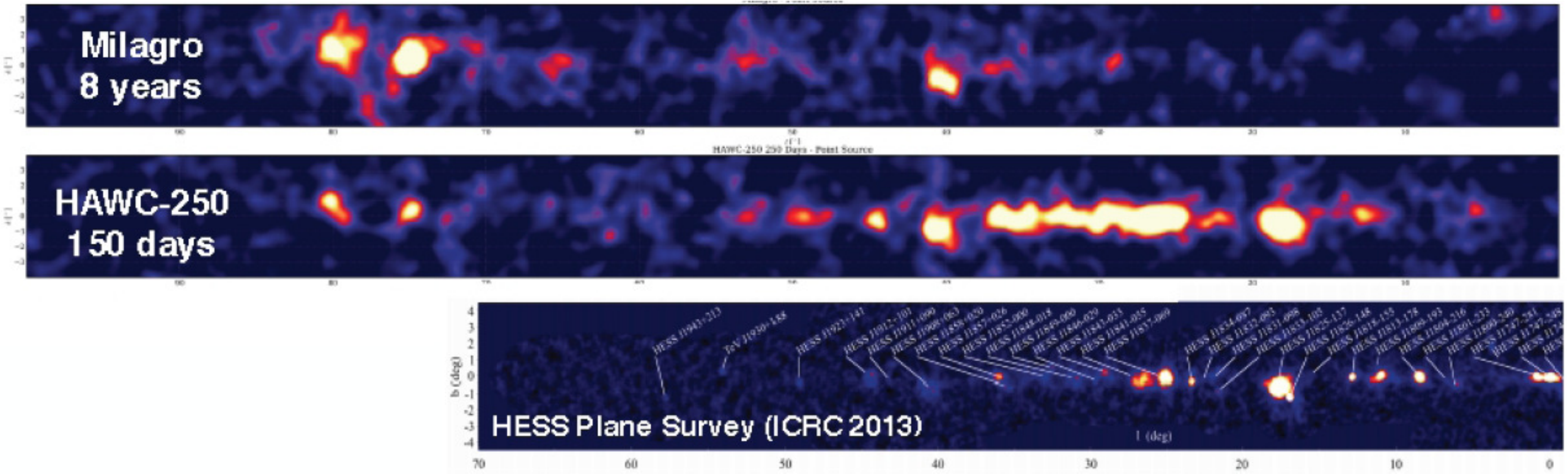}
  \caption{Comparison of the significance maps of the inner Galaxy region for 8 years of Milagro data (Top), 150 days of HAWC-250 data (Middle), the H.E.S.S. Galactic Plane survey presented at the last ICRC (Bottom).
}
  \label{fig:hawc_plane}
\end{figure}

With its large field of view and location at 19$^{\circ}$ N latitude, HAWC is surveying the Galactic Plane from high Galactic longitudes down to near the Galactic Center. Due to the limited angular resolution of HAWC, a major challenge analyzing the emission from the inner Galaxy region where source confusion is frequent, is to deconvolute and identify sources. To do so, a likelihood framework has been developed to simultaneously fit the positions and fluxes of multiple sources, allowing to determine the number of sources in a region of interest (ROI). During this ICRC, an analysis of the inner Galaxy region of l $\in$ [+15$^{\circ}$, +50$^{\circ}$] and b $\in$ [--4$^{\circ}$, +4$^{\circ}$] using the data taken with a partially-completed HAWC array has been presented, showing ten source candidates identified with $> 3\sigma$ post-trials, eight of them having tentative TeV associations~\cite{hawc_plane}. The high sensitivity of HAWC in comparison to Milagro for such study and the consistency with the H.E.S.S. observations on the same region is very well highlighted in Figure~\ref{fig:hawc_plane}. Most sources are either pulsar wind nebula (PWN) candidates or unidentified sources (UNID). In the end, it is very likely that a dominant fraction of the sources detected by HAWC will be PWNe, as it is already the case with TeV telescopes.

\subsection{Pulsar wind nebulae}
The number of TeV sources in this class is still growing and the recent detection by MAGIC of the PWN 3C58 is a very good example~\cite{3C58}. This PWN is the faintest detected at TeV energies and the least energetic. Thanks to the large number of PWNe detected by TeV experiments, populations studies are now being carried out. Interestingly, using a simple modeling, it is possible to describe the trends and scatter of the present PWN population. For instance, all PWNe candidates detected at TeV energies are powered by relatively young pulsars with high spin down power. However, variations between the different systems are also clearly visible and could be due to the differences in the surrounding medium or different starting conditions~\cite{hess_pwnpop}.

We cannot present PWNe without saying a word on the most famous one: the Crab Nebula. It is now well known that strong flares have been detected at GeV energies by AGILE and \emph{Fermi}-LAT. At TeV energies, the Crab nebula seems to be much more quiet and no variability was detected so far by the different experiments. Being one of the best studied objects, the precision of the measurements obtained by the different experiments is now extremely impressive. For instance, the spectral measurements obtained by the MAGIC collaboration cover more than three decades in energy~\cite{magic_crab}. Modeling these data is now a real challenge :  simple models (constant magnetic field or spherical symmetry) are unable to reproduce the current TeV data. On the same source, the H.E.S.S. collaboration successfully presented results from the full hybrid H.E.S.S. array, applying a method which combines monoscopic and stereoscopic events into one overall analysis~\cite{hess_crab}. These results are important in two respects: this lowers the energy threshold and increases the statistics to constrain even further theoretical models on PWNe, and it offers groundwork for the future Cherenkov Telescope Array (CTA) which will also be a hybrid system. Finally, still on the Crab nebula, search for continuous gamma-ray emission above 100 TeV has been performed by the Tibet AS array~\cite{tibet_crab}. No significant excess was found, and the upper limit obtained above 140 TeV is most constraining until now.

Geminga is an important high energy source. With its location at approximately 250 pc from us, it is one of the closest known middle aged pulsars. The relative proximity of Geminga raises an interesting possibility, namely that the high-energy particles accelerated by the PWN, most likely electrons and positrons, may be at the root of the explanation of the "positron excess". And indeed, some authors \cite{positron} have used the detection of an extended source spatially consistent with Geminga by Milagro and have shown that this source could be the evidence for the production, acceleration and escape of electrons with energies up to 100 TeV.  HAWC-250 also sees an excess of 6.3$\sigma$ at the location of the Geminga pulsar using an additional 3$^{\circ}$ smoothing~\cite{hawc_geminga}. This new result therefore confirms Milagro's observation. In addition, the slightly lower significance obtained by HAWC in comparison to Milagro, despite a larger dataset, favors a spectrum harder than the Crab although a Crab-like spectrum cannot be ruled out. It should be noted that no detection was reported by MAGIC and VERITAS despite intense observations but such large sources are extremely hard to detect with small field of view IACTs~\cite{veritas_geminga, magic_geminga}. Two analysis techniques are being implemented by VERITAS which will help in the study of spatially extended gamma-ray emission which may emanate from this region. These techniques are progressing well and should be ready to analyze the VERITAS Geminga data in the near future.

This directly leads us to the topic of pulsars which are powering these bright and numerous PWNe. 

\subsection{Pulsars}
\begin{figure}[t]
 \centering
     \includegraphics[height=0.35\textheight]{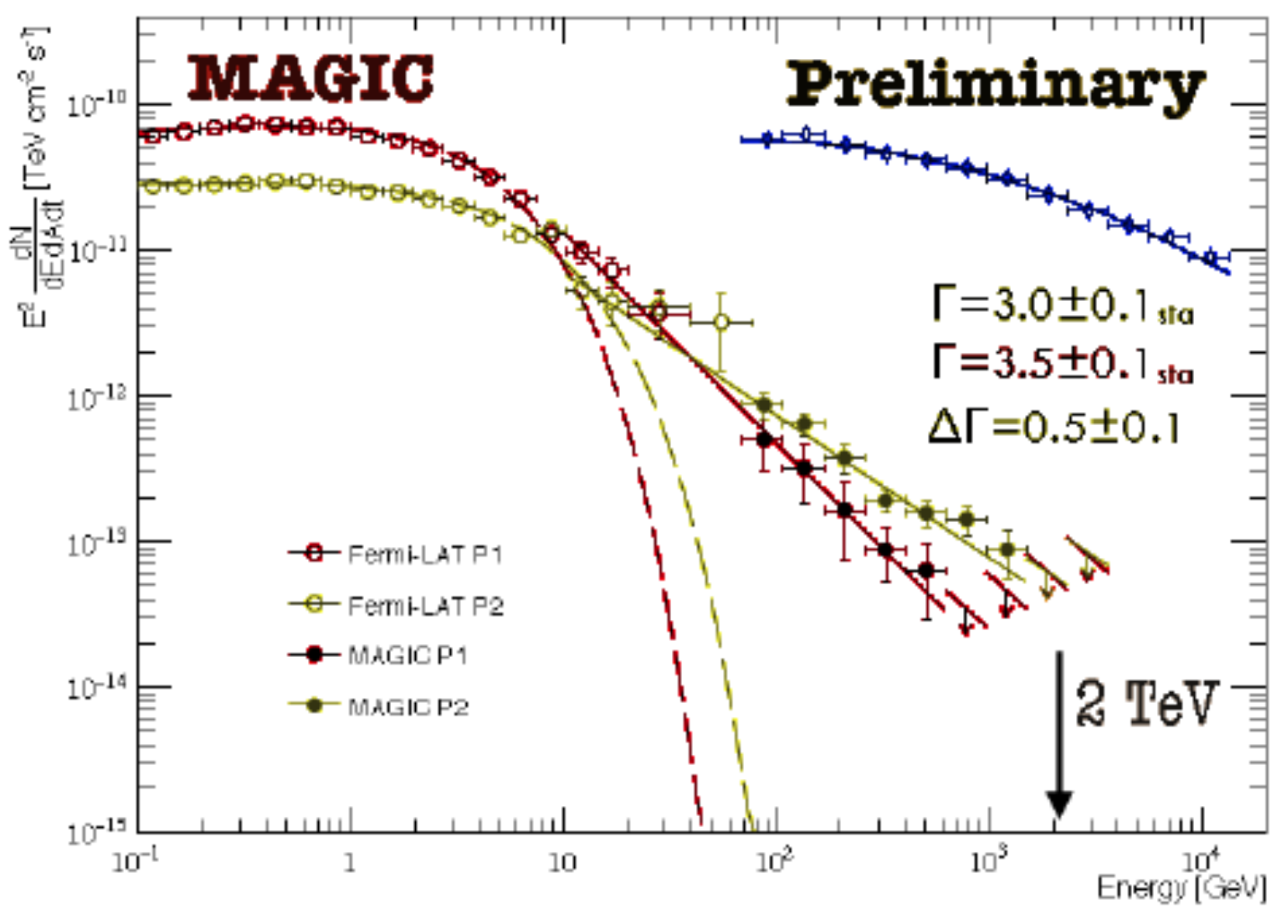}
  \caption{Spectral  energy  distributions  of  the  two peaks P1 and P2 of the Crab pulsed emission as measured with
\emph{Fermi}-LAT (below 100 GeV, open circles) and MAGIC (above 70 GeV, filled circles)~\cite{magic_crab}. The MAGIC flux of the Crab nebula is also plotted in blue for comparison.
}
  \label{fig:crab}
\end{figure}

The search for pulsed emission from the Crab pulsar in very-high energy gamma rays has a long history. The MAGIC telescope was the first to detect pulsed emission above 25 GeV~\cite{crabmagic_science}. Some time later, observations with VERITAS revealed pulsed gamma-ray emissions at energies above 100 GeV~\cite{crabveritas_science}. During this ICRC, VERITAS presented an updated energy spectrum that extends beyond 400 GeV~\cite{veritas_crab}. On its side, still on the Crab, the MAGIC collaboration is now seeing significant pulsed emission above 1 TeV as can be seen Figure in \ref{fig:crab}~\cite{magic_crab}. These new results are a real challenge for pulsar models. It is impossible to reach such high energies with synchro-curvature which would support the inverse Compton mechanism. In addition, to avoid fast cooling due to the large magnetic fields in the magnetosphere, the emission region must be far from the surface of the neutron star, in the vicinity of the termination of the magnetosphere. However, keeping a pulsed signal through inverse Compton radiation at such large distance from the pulsar is not trivial. Being one of the best studied objects, the Crab is still providing new surprises.\\
A second detection of pulsed emission from a pulsar, Vela, was recently announced by H.E.S.S. down to energies of 20 GeV thanks to the use of H.E.S.S.II data in monoscopic mode~\cite{hess_vela}. However, a deep observation campaign will be needed to investigate and constrain the maximum energy of the detected gamma-ray photons. 

\subsection{Gamma-ray binaries}
Pulsars are powering PWNe but they also can play the role of compact object in gamma-ray binaries such as PSR B1259$-$63 that represents the only case in this specific class of systems for which the nature of the compact object, a neutron star with a spin period of 48 ms, has been established. Although the new H.E.S.S.II data does not provide any significant evidence of an abrupt increase of the VHE emission similar to the HE case with \emph{Fermi}-LAT, the new 2014 observations nevertheless reveal a relatively high source flux state at TeV energies $\sim$50 days after periastron~\cite{hess_psrb1259}. The case of LS I +61$^{\circ}$ 303 is extremely intriguing. The flux from this gamma-ray binary varies strongly with the orbital period of 26.5 days. The maximum VHE flux is found around apastron at a level typically corresponding to 10 -- 15\% of the Crab Nebula flux (>300 GeV). During recent VERITAS observations, relatively short (day scale), bright TeV flares were observed from the source around apastron in two orbital cycles (October and November 2014). Both cases exhibited peak fluxes above 25\% of the Crab nebula flux (>300 GeV), making these the brightest VHE flares ever detected from this source. Different scenarii are proposed to understand the origin of these flares, depending on the nature of the compact object: a microquasar scenario or a pulsar binary scenario. Further multi-wavelength observations of LS I +61$^{\circ}$ 303 are therefore necessary to fully understand the varying TeV emission from the source and determine the nature of the compact object.

\subsection{Supernova remnants}
\begin{figure}[t]
 \centering
     \includegraphics[height=0.25\textheight]{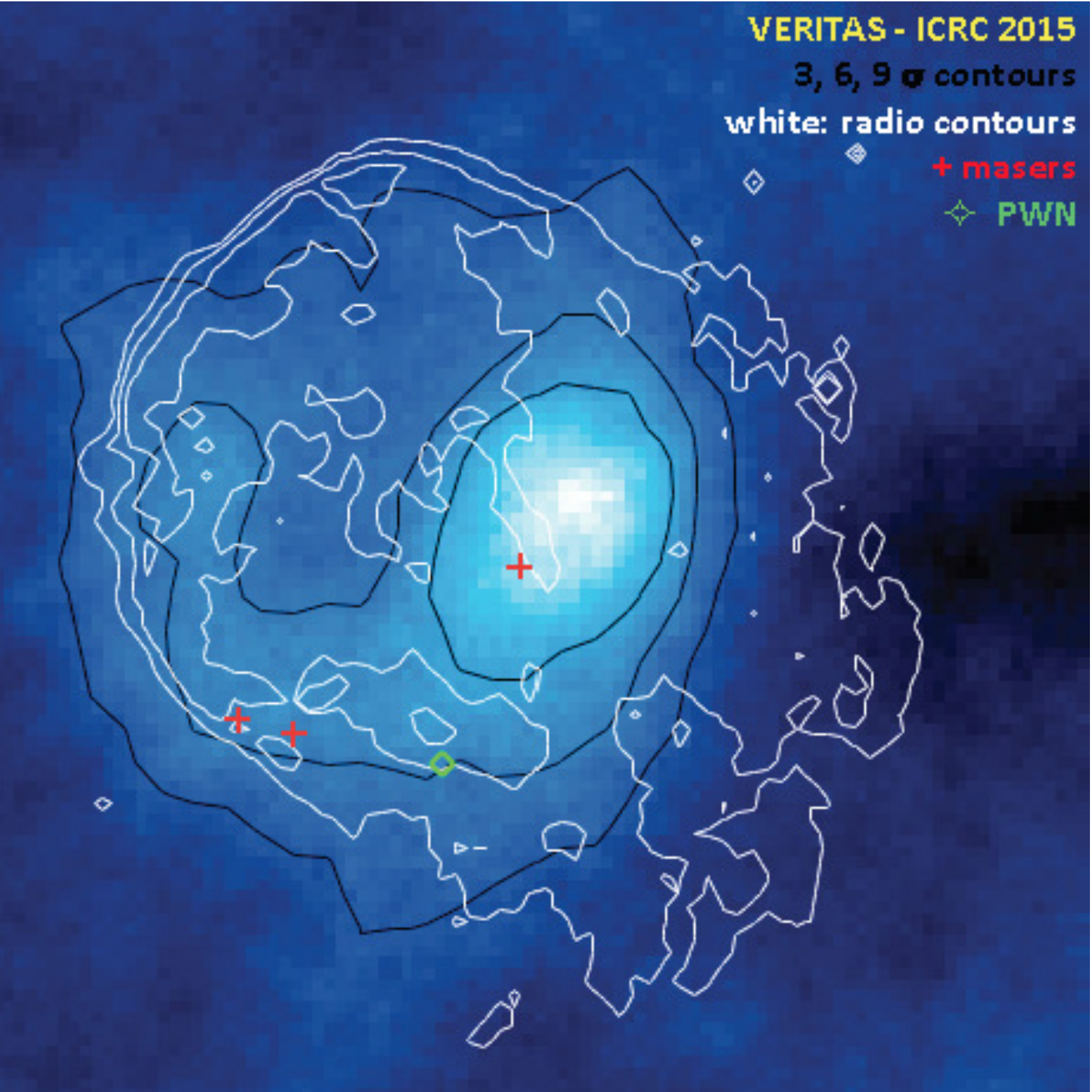}
	\includegraphics[height=0.25\textheight]{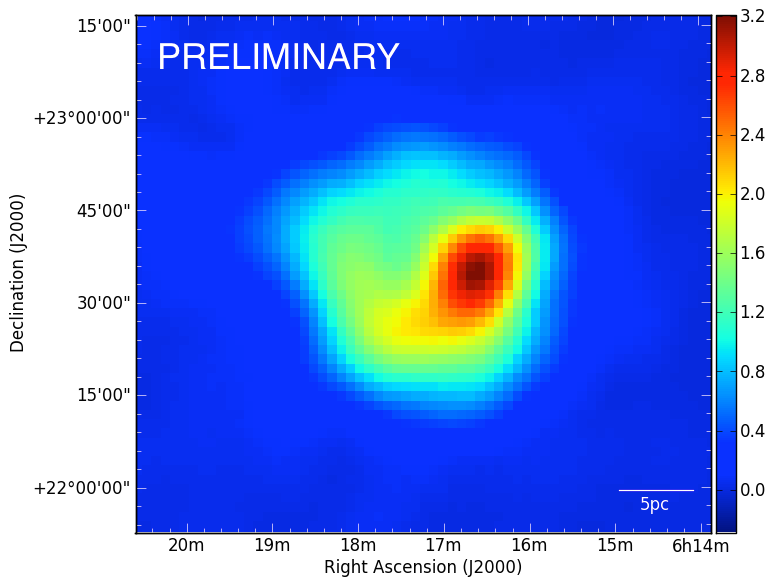}
  \caption{Left: Gamma-ray excess map of IC 443 as viewed by VERITAS. Black contours are at 3, 6, 9$\sigma$ level. Locations of OH maser emission are represented by the red crosses while the position of the PWN CXOU J061705.3+222127 is indicated by a green diamond. White contours show the radio emission. Right: Counts map of the region of IC443 using 83 months of \emph{Fermi}-LAT Pass 8 data, PSF2+PSF3 events from 5 to 500 GeV. The image has been smoothed with a Gaussian kernal radius of 9 arcminutes.
}
  \label{fig:ic443}
\end{figure}

Supernova remnant represent another important type of objects in our Galaxy and one of the prime candidates for the acceleration of cosmic rays. 
However, the identification of TeV sources as SNRs is absolutely not trivial and entirely relies on their shell-like appearance and a TeV morphology matching their shell-like counterparts in radio and non-thermal X-rays. Using the increased data set of the Galactic Plane Survey, the HESS collaboration has reported the detection and identification of one new shell-type SNR candidate~\cite{hess_snrs}, HESS J1534$-$571, is coincident with the radio SNR G323.7$-$1.0 recently detected in MGPS2 data~\cite{g323}. This demonstrates the capability of the current generation of TeV instruments to discover new SNRs. Increasing the data set can also allow to better understand already known objects, such as Tycho and RX J1713.7$-$3946. In the case of Tycho, the new and softer spectrum, down to 400 GeV thanks to the improved low energy sensitivity and deeper exposure, reported by the VERITAS collaboration is now in clear tension with previous models~\cite{tycho}. These new data may help to constrain the maximum energy of the particles accelerated in this remnant. For what concern RX~J1713.7$-$3946, the angular resolution, better than 0.05 deg, and the increased dataset of 150 hours of observation enable for the first time a detailed investigation of morphological differences between TeV gamma rays and X-rays. The broader radial profile seen in gamma-rays in comparison to X-rays may be the first evidence of escape of protons that would emit gamma rays through proton-proton interactions when interacting with surrounding matter. Alternatively, regions of low magnetic field values could also be bright in gamma-rays while being rather faint in X-rays, explaining as well the differences between the gamma-ray and X-ray detected signals~\cite{rxj}. The case of IC 443 is one of the most exciting results on SNRs at this ICRC. The deep observation campaign performed by VERITAS allows the first morphological study of this middle-age remnant in interaction with a molecular cloud, revealing a bright gamma-ray emission at the position of the brightest maser but also some emission throughout the entire northeast lobe~\cite{ic443}. Interestingly \emph{Fermi}-LAT sees a very similar morphology above 5 GeV, as shown in Figure~\ref{fig:ic443}. The spectral analysis that is going on in different regions of the remnant seems to show a harder spectrum in the North East (where there is no interaction with the molecular cloud) which is extremely important to constrain and probe the environmental dependence of cosmic-ray diffusion. 

\subsection{The Galactic center and the diffuse emission from the Galactic plane}
\label{gc}
\begin{figure}[t]
 \centering
     \includegraphics[height=0.3\textheight]{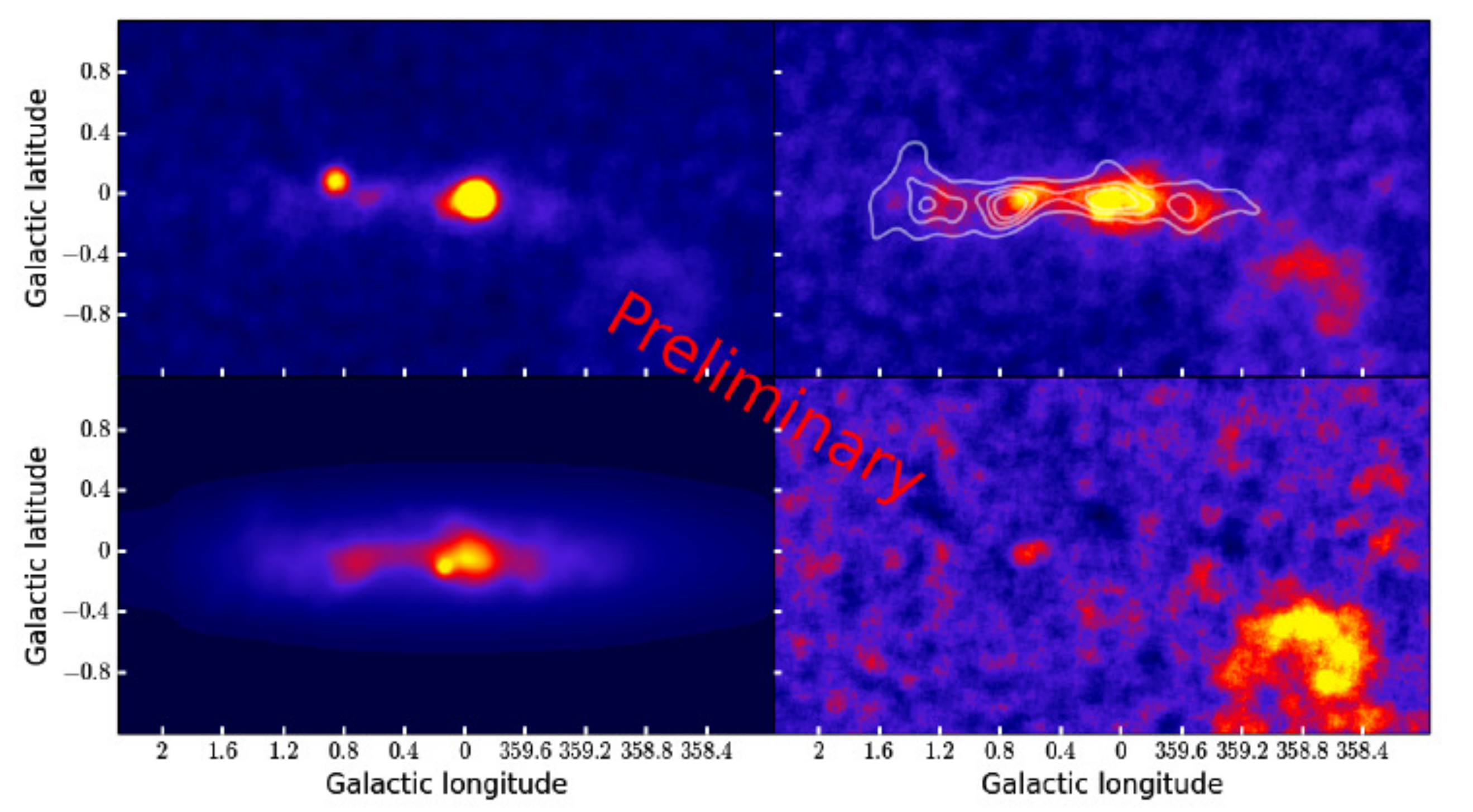}
  \caption{Top Panels: New VHE g-ray images of the GC region as seen by H.E.S.S., in Galactic coordinates and smoothed with the H.E.S.S. PSF~\cite{gc_diffuse}. Left panel: gamma-ray significance map. Right panel: same map after subtraction of the two point sources G0.9+0.1 and HESS J1745$-$290. The white contours indicate the density of molecular gas as traced by the CS emission and smoothed with the H.E.S.S PSF. Bottom left panel: best model of the GC Ridge VHE emission obtained with this study. Bottom right panel: map of the residual significances after complete subtraction of the model described in Section~2.6. 
  }
  \label{fig:gc}
\end{figure}

The Galactic center (GC) region also hosts several potential cosmic-ray accelerators. Indeed, the hard spectrum of the ridge emission, revealed by H.E.S.S. in 2006, and its spatial correlation with the local gas density suggest that the emission detected was due to collisions of multi-TeV cosmic rays with the dense clouds of interstellar gas present in this region. Now thanks to a total of 260 hours, the H.E.S.S. collaboration has performed a thorough analysis showing that $\sim$half of GC ridge emission is distributed like dense gas tracers over a projected distance of 140 pc and fades beyond~\cite{gc_diffuse}. An additional large scale emission that does not correlate with dense gas tracers, and could be the result of unresolved sources and/or a gas component in a diffuse phase not seen by gas tracers, is also required as can be seen in Figure~\ref{fig:gc}. Finally, a new source is detected, dubbed HESS J1746$-$285, spatially coincident with the Fermi Arc source and the X-ray PWN candidate G0.13$-$0.13. This source is also detected by VERITAS and MAGIC at similar positions, with a general diffuse emission in excellent agreement with the observation made by H.E.S.S.~\cite{gc_veritas, gc_magic}. These complementary observations are extremely useful since the two Northern experiments are seeing the GC at larger zenith angle and therefore probe higher energy particles.\\ 
In addition to the three components listed above, a central emission extending over $\sim$15 pc is detected. It could be the signature of a radial gradient of CRs in the central molecular zone (CMZ) that is expected if they are accelerated by the SMBH itself. Interestingly, the measured CR density profile in this region appears to be in a quite good agreement with a 1/r profile. It is remarkable that such a dependence is expected in the case of an accelerator located in the inner 10 pc region of GC continuously injecting particles for more than $10^4$ years. Finally, the energy spectrum of the diffuse gamma-ray emission, extracted from a ring centered at the GC up to a radius of 60 pc is well described by a pure power-law with a photon index 2.3 up to 40 TeV, without any indication of energy cut-off or break. The hadronic origin of the diffuse VHE emission implies that gamma rays result from the decay of neutral pions produced by relativistic protons (at least 2.8 PeV at 68\% confidence level) interacting with the interstellar gas. This is the first robust identification of a VHE cosmic hadronic accelerator operating as a PeVatron~\cite{gc_hess} !

At higher energy (E $>$ 350 GeV) and longitudes comprised between 25$^{\circ}$ and 100$^{\circ}$, more than five years of ARGO-YBJ data have been used to study the diffuse gamma-rays from the Galactic plane. A spectral analysis has been carried out, showing an energy spectrum  slightly softer than that of the \emph{Fermi}-LAT Galactic diffuse emission model but consistent at 1$\sigma$ level. On the other hand, the TeV flux averaged over the Cygnus region 65$^{\circ}$ $<$ l $<$ 85$^{\circ}$ shows a marginal evidence of a harder spectrum, indicating the possible presence of young cosmic rays coming from a nearby source~\cite{argo_diffuse}. More data, especially with HAWC, would be extremely useful to confirm this result.

\subsection{The Large Magellanic Cloud}
As a final point on Galactic sources and moving to the extragalactic sky, the Large Magellanic Cloud, a dwarf satellite galaxy of our Milky Way, is extremely instructive and encouraging for the future. Indeed, using a total of 210  hours, H.E.S.S. detected  three extremely energetic objects of different  type within the LMC~\cite{lmc}: 
\begin{itemize}
\item the PWN N~157B, a counterpart to the Crab nebula but with an electron acceleration efficiency is five times lower,
\item the SNR N~132D, one of the oldest VHE gamma-ray emitting SNRs, 
\item the superbubble 30 Dor C, thus putting into light a new class of VHE emitters. 
\end{itemize}
It is the first  time in a galaxy outside the Milky Way, that individual sources of very high energy gamma-rays can be resolved. Interestingly, the unique object SN 1987A is, surprisingly, not detected, which constrains the theoretical framework of particle acceleration
in very young supernova remnants. These results open a new window at TeV energies and we can expect further discoveries with more sensitive surveys of the LMC in gamma-rays, for instance with the Cherenkov Telescope Array.

\section{The extragalactic sky}
The two last years since the previous ICRC have been very active for gamma-ray blazars, AGNs with jets pointing close to the line of sight, resulting in several new TeV detections of very interesting objects: S3 0218+357, PKS 1441+25, RGB J2243+203, and S3 1227+25. With these new detections, blazars are now detected at VHE out to an extreme redshift of z = 0.944, while until recently the farthest sources observed in this energy range were 3C 279 (z = 0.536), KUV 00311-1938 (z > 0.506) and PKS1424+240 (z > 0.6). 

\subsection{The Extragalactic background light}
\begin{figure}[h]
 \centering
     \includegraphics[height=0.3\textheight]{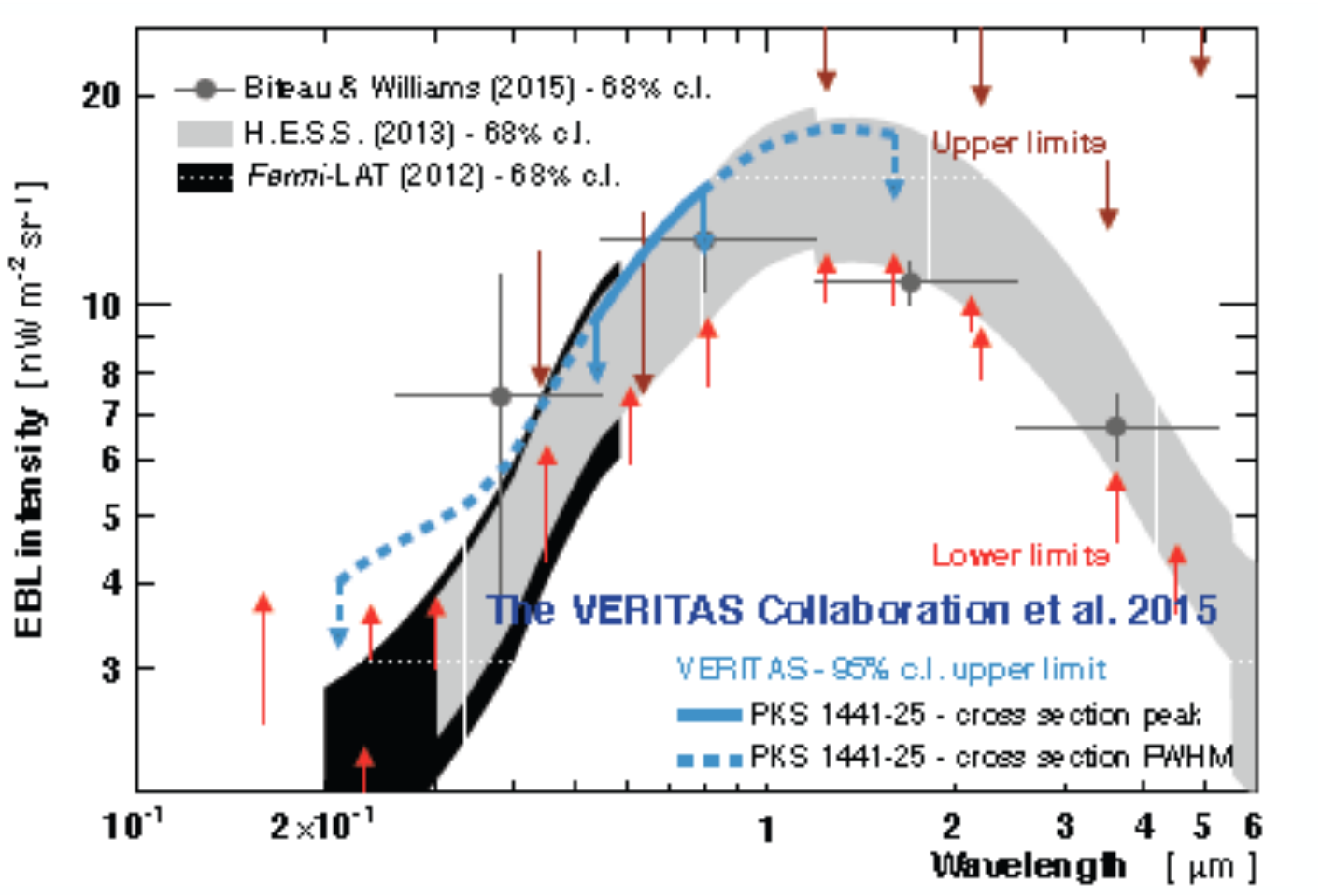}
  \caption{Specific intensity of the EBL. Galaxy counts and direct measurements are extracted from the literature and represented by upward- and downward-going arrows, respectively. The model-independent result by \cite{ebl} is shown with gray points, while the model-dependent results based on the EBL intensity by \cite{ebl_hess} and \cite{ebl_fermi} are shown with filled gray and black regions, respectively (1$\sigma$ confidence level).
}
  \label{fig:ebl}
\end{figure}

Observations of farther sources in VHE gamma-rays are difficult due to the strong absorption in the interaction with the background radiation field originating from starlight emission and its re-processing by interstellar medium integrated over cosmic history, called the extragalactic background light (EBL). VHE gamma-rays interact with IR to UV photons via electron-positron pair production, resulting in an attenuated observed flux and effectively creating a gamma-ray "horizon". In this context, the recent detection of PKS 1441+25 by MAGIC is extremely interesting since it is the most distant blazar detected by TeV experiments with its redshift of z=0.939~\cite{pks1441_magic}. Immediately after the MAGIC discovery in April 2015, VERITAS initiated a ToO observation campaign. A total of 15 hours of good-quality observations was acquired during a 1-week period, resulting in the detection of a very soft spectrum excess of $\sim$400 events~\cite{pks1441_veritas}. Thanks to its high redshift and using the spectrum derived from these observations, preliminary Extragalactic Background Light (EBL) constraints for this flare are highly competitive in comparison to previous measurements using a sample of sources, as can be seen in Figure~\ref{fig:ebl}. 

\subsection{Gravitational lensing}
It is clear from the example of PKS 1441+25 that, to maximize the detection chance of high redshift blazar, the observations are often triggered by a high state observed in lower energy ranges. In particular, \emph{Fermi}-LAT scanning the whole sky in GeV range can provide alerts of high energy fluxes and spectral shape. It was the case for QSO B0218+357, also known as S3 0218+35, a blazar located at the redshift of 0.944. QSO B0218+357 is one of only two objects with a measured gravitational lensing effect in GeV energy range. In 2012, it went through a series of outbursts registered by the \emph{Fermi}-LAT instrument. The statistical analysis of the light curve autocorrelation function led to a measurement of time delay between the direct and the lensed components of 11.46 $\pm$ 0.16 days. Another flaring state of QSO B0218+357 was observed by \emph{Fermi}-LAT in July 2014. Unfortunately, the first component was not visible by MAGIC due to the full moon period but they could catch the second (lensed) component. It is therefore the first gravitationally lensed blazar detected at the VHE energies. Interestingly, the emission is very bright despite the attenuation at such redshift which can be used to put constraints on the EBL. 

\subsection{Multi-wavelength campaign}
The blazar Markarian 501 is among the brightest X-ray and TeV sources in the sky, and among the few sources whose radio to VHE spectral energy distributions can be characterized by current instruments by means of relatively short observations (minutes to hours). In 2013, an extensive multi-instrument campaign involving the participation of \emph{Fermi}-LAT, MAGIC, VERITAS, F-GAMMA, \emph{Swift}, GASP-WEBT, and other collaborations/groups and instruments was organized~\cite{mkn501}. It provided the most detailed temporal and energy coverage on Mrk 501 to date. This observing campaign included, for the first time, observations with the Nuclear Stereoscopic Telescope Array (NuSTAR), which is a satellite mission launched in June 2012. NuSTAR provides unprecedented sensitivity in the hard X-ray range 3 -- 79 keV, which, together with MAGIC and VERITAS observations, is crucial to probe the highest energy electrons in Mrk 501. A significant correlation between the X-ray fluxes (NuSTAR and Swift/XRT) and the VHE emissions (MAGIC and VERITAS) was detected. Interestingly, a large fraction of the MAGIC data were affected by sand from the Saharan desert, in particular during the flaring activity. These data have been corrected using the atmospheric information from the LIDAR facility that is operational at the MAGIC site on an event by event basis. It is the first time that LIDAR information is used to produce a physics result with Cherenkov telescope data taken during adverse atmospheric conditions. 

Obviously, with an instantaneous field of view of 2 sr, particle detectors like ARGO-YBJ and HAWC can survey two-thirds of the sky every day. These unprecedented observational capabilities allow to continuously scan the highly variable extragalactic gamma-ray sky providing complementary measurements during a MWL campaign.  Multi-wavelength observations of Mrk 421 over 4.5 years, from 2008 August to 2013 February, were presented by the ARGO-YBJ collaboration~\cite{mkn421_argo}. According to the observed light curves, ten states (including seven large flares, two quiescent phases and one outburst) were selected and systematically analyzed. The underlying physical mechanisms responsible for different states may be related to the acceleration process or to variations of the ambient medium. ARGO-YBJ is no longer operational but the deployment of HAWC is now complete and first flux light curves, binned in week-long intervals, were presented at this ICRC  for the TeV-emitting blazars Markarian 421 and 501~\cite{blazars_hawc}. On both sources, indications of gamma-ray flare observations were shown, demonstrating that a water Cherenkov detector can monitor TeV-scale variability of extragalactic sources on weekly time scales.

\subsection{Radio galaxies}
\begin{figure}[t]
 \centering
     \includegraphics[height=0.3\textheight]{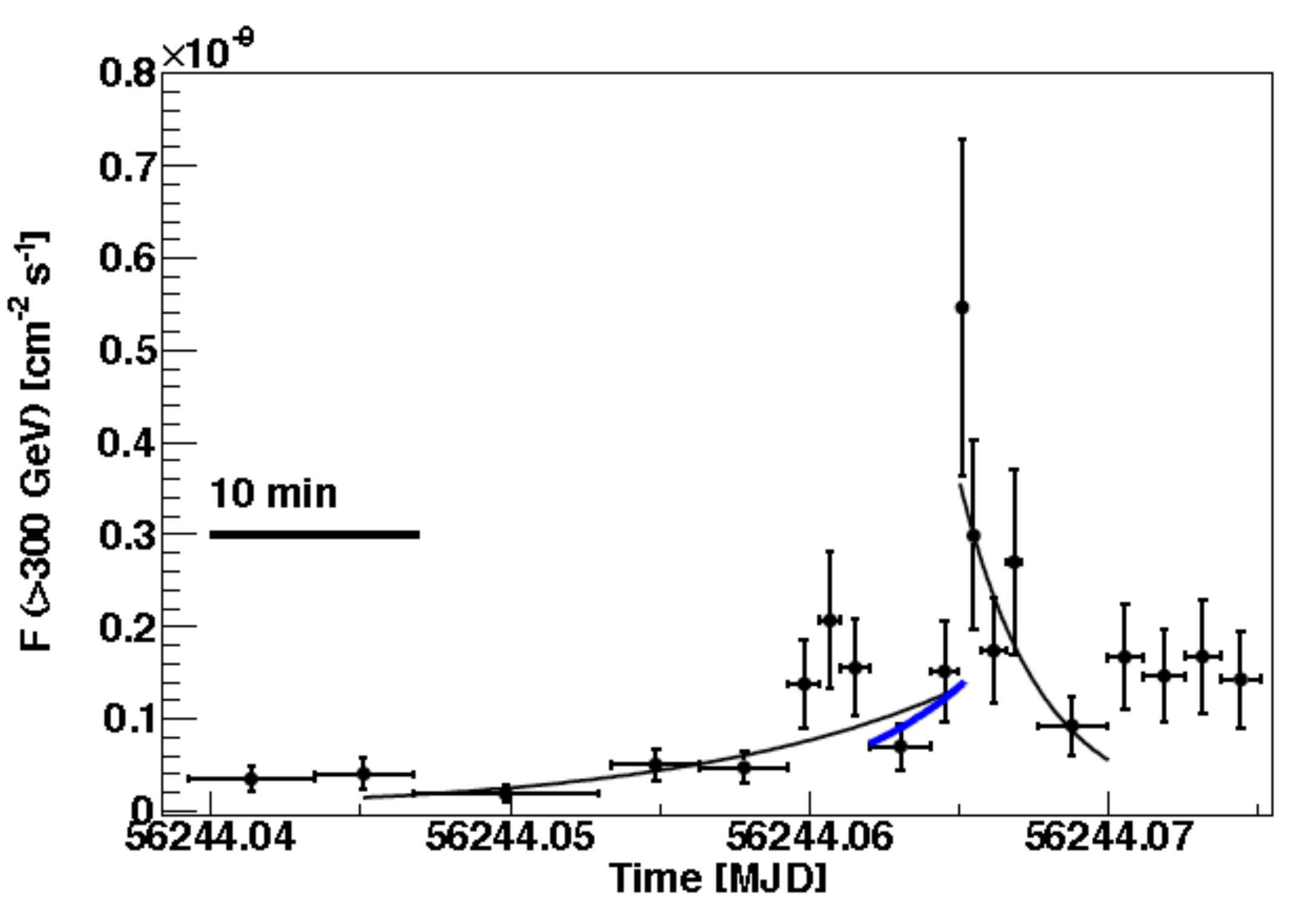}
  \caption{Zoom of the MAGIC light curve of the flare of IC~310 calculated above 300~GeV into the time range 00:57
(MJD 56244.04) to 01:40 (MJD 56244.07)~\cite{ic310}. The black lines show exponential fits to the light curve to the
rising and decay phases of the peak. The blue line shows the fit corresponding to the slowest doubling time
necessary to explain the rising part of the flare at confidence level of 95\%. 
}
  \label{fig:ic310}
\end{figure}

Blazars, which have been discussed in detail above, are the most numerous class of extragalactic objects discovered at VHE. However, there is a growing evidence that blazars are not the only extragalactic objects capable of VHE emission. With the detection of M 87, Cen A, IC 310, Per A and PKS 0625-354~\cite{pks0625} nearby radio galaxies (RGs) seem to constitute a new class of VHE sources. RGs are active galaxies with their relativistic jets oriented at intermediate to larger viewing angles with respect to the line of sight. As a result of larger inclinations, the observed non-thermal emission produced within the innermost parts of the jets is not amplified by relativistic beaming and hence different emission components, typically not present in observed blazar spectra, may become prominent. Therefore, since RGs are considered as blazars observed at larger viewing angles, modeling of such sources provides an independent check of blazar models. In addition, gamma-ray observations of RGs may reveal some non-standard processes possibly related to the production of high energy photons and particles within active nuclei and extended lobes. And third, increasing the sample of gamma-ray RGs will enable to understand the contribution of nearby non-blazar AGN to the extragalactic gamma-ray background. These sources are therefore of very high interest and the example of IC~310 is excellent in this respect. 
Using radio data, the angle of the jet could be constrained between 10 and 20$^{\circ}$ which means that this is not a blazar. For the range of orientation angles inferred from radio observations, the Doppler factor is constrained to a value smaller than 6. However, the MAGIC collaboration detected very strong variability from this source with a flux doubling time of 4.88 min for the rising phase (see Figure~\ref{fig:ic310}). The variability of 4.8 min constrains the radius of the emission region to be $\sim20$\% of the event horizon: a very high value of the Doppler factor is thus required to avoid the absorption of the gamma rays due to interactions with low-energy synchrotron photons. In summary, trying to interpret the data in the frame of the shock-in-jet model meets difficulties. The MAGIC collaboration suggests that the emission would be associated with pulsar-like particle acceleration by the electric field across a magnetospheric gap at the base of the radio jet~\cite{ic310}. The emission would then be produced by either inverse-Compton scattering on a background photon field, or by curvature radiation. To finally answer which kind of mechanism is responsible for ultra-rapid flux variability events, more observations are needed, e.g., with higher sensitivity as provided by the Cherenkov Telescope Array as this potentially allows to measure even faster flares as well as rapid flux variations from other, e.g., fainter AGN. Furthermore observing such events with simultaneous multi-wavelength coverage can help to constrain emission models.

\subsection{Gamma-ray bursts}
Gamma-ray bursts (GRBs) are the most luminous, highly-relativistic light sources known and may be generated during the collapse of a massive star (long GRBs) or via a merger event (short GRBs). They emit light across the electromagnetic spectrum, including at GeV energies and hence provide key targets for VHE gamma-ray detectors. And, indeed, a recent observation of the \emph{Fermi}-LAT has established that GRBs produce photons in the very-high-energy (VHE, > 100 GeV) regime, when a 95.3 GeV photon (or 128 GeV when corrected for redshift) was detected in GRB 130427A~\cite{monstergrb}. As the gamma-ray luminosity of GRBs falls off very rapidly after burst time, in order to maximize the chances of a GRB detection at VHE it is crucial to begin observations as soon as possible. With an instantaneous field of view of approximately 2 sr and over 95\% duty cycle (up time fraction), HAWC is an ideal detector to perform ground-based gamma-ray observations of GRBs. Though optimized for TeV observations, HAWC has significant sensitivity to short transients of energies as small as 50 GeV. It was shown that a GRB similar to 130427A would be detected even if the spectrum doesn't extend beyond what was observed by the LAT~\cite{monster_hawc}. In addition to exceptional bursts, HAWC might detect other GRBs with a rate as high as 1--2 GRBs per year. At this ICRC, initial results for 18 \emph{Swift}-detected GRBs in the HAWC field of view during HAWC-111 were shown. None of the GRBs is significant above 3$\sigma$ when accounting for trial factors~\cite{hawc_grb}.
On the other side, the lower energy threshold of IACTs such as MAGIC and HESS is also an excellent advantage for the detection of GRBs, since it reduces the flux attenuation by pair production with the lower energy (optical/IR) photons of the EBL. Due to their limited field of view, IACTs must be equipped with a very fast repointing of the telescopes. For the case of H.E.S.S., in order to minimize this delay, two major improvements have been made for the additional telescope (CT5) over the original 4 telescopes. Firstly the telescope drive system of CT5 is significantly updated over that of the original H.E.S.S. system, and is able to perform a full rotation of the telescope (180$^{\circ}$ in azimuth) in $\sim$110 seconds. Additionally CT5 is permitted to point in reverse-mode, allowing the telescope to slew through zenith, resulting in significantly faster repointing for some GRBs, where otherwise a large azimuthal slew would be required. In addition to this rapid slewing a fully automatic target of opportunity (ToO) observation system has been implemented within the H.E.S.S. and MAGIC arrays, in this case receiving triggers from the GCN system \cite{hess_grb, magic_grb}. In the last two years, the upgrade of the MAGIC system and an improved GRB observation procedure has made possible follow-up of GRBs within 100s after the event onset. The preliminary data analysis of 13 GRBs observed with this improved automatic procedure did not allow the detection of any significant signal, however, these new developments open a new phase in the study of GRBs. 

\section{The future}
As discussed in Section~\ref{status}, major upgrades to all of the current generation of IACTs have recently been completed. For the future, new facilities are now in the prototyping stage and are developed to offer more statistics, better quality of photon reconstruction and new "type" of photons reconstructed (going to the lower or higher energy range).

\subsection{The Cherenkov Telescope Array}
\begin{figure}[t]
 \centering
     \includegraphics[height=0.35\textheight]{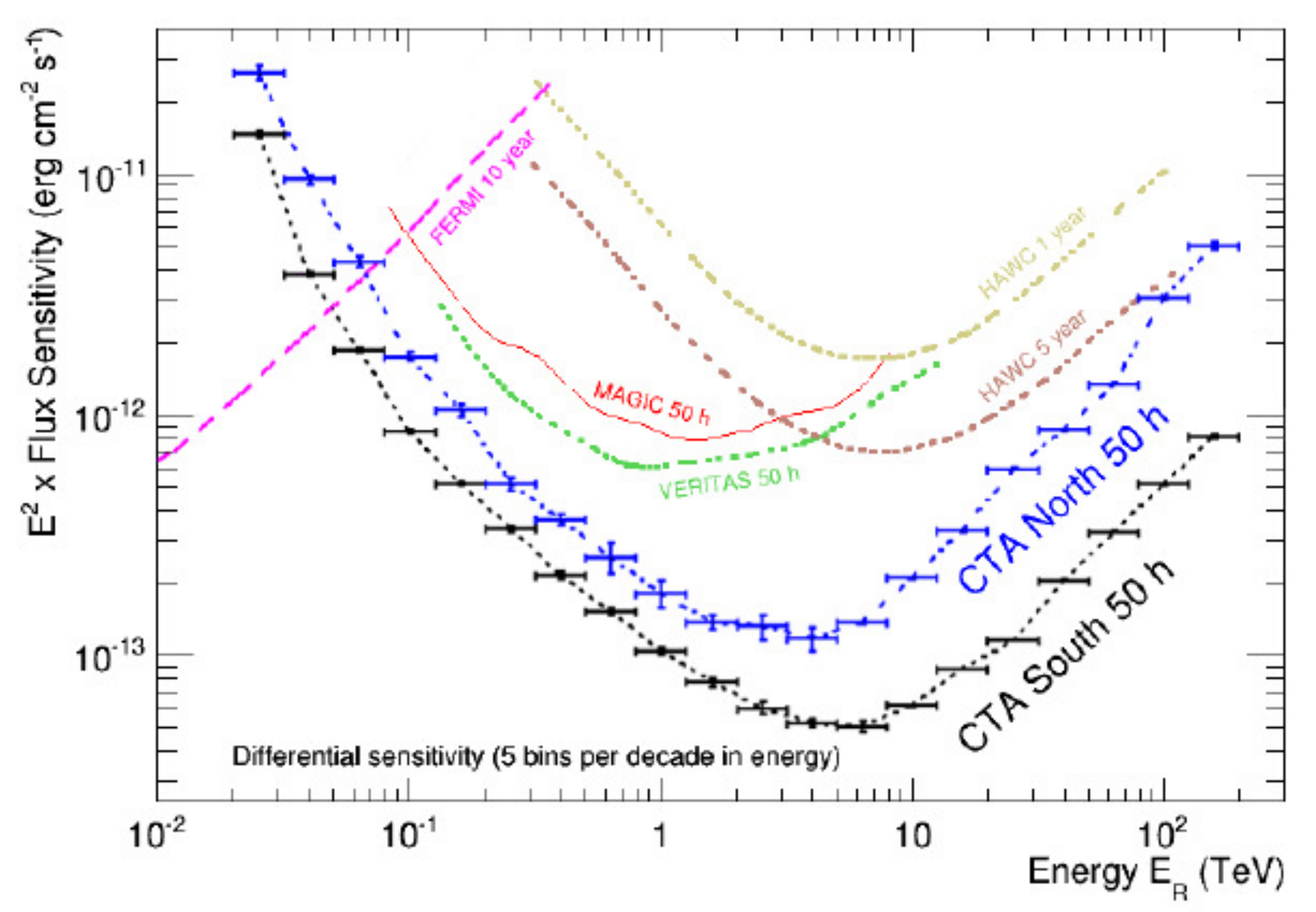}
  \caption{Differential sensitivity for a point-like g-ray source of the CTA-N and CTA-S candidate arrays (50 hours of observation, N/S pointing average) \cite{cta_sensitivity} together with the current MAGIC and VERITAS (50 hours) and and future Fermi-LAT (over 10 years of operation) and HAWC (1 and 5 years) attained sensitivities.
}
  \label{fig:cta}
\end{figure}

One of the best solutions covering these 3 requirements is the Cherenkov Telescope Array (CTA). CTA is an array of about 50--100 Cherenkov telescopes per site at two sites in the southern and the northern hemispheres, thus allowing full-sky coverage. At the time of writing, formal negotiations on two sites (Paranal in Chile and la Palma in the Canaries Islands in Spain) are underway. The large number of telescopes will come in three size classes: Large Size Telescopes (LSTs) sensitive to the low energy showers (below 200 GeV)~\cite{cta_lst}, Medium Size Telescopes (MSTs) increasing the effective area within the CTA core energy range (between 100 GeV and 10 TeV)~\cite{cta_mst} and Small Size Telescopes (SSTs, for CTA South only)~\cite{cta_sst} spread out over several km$^2$ to catch the rare events at the highest energies of the electromagnetic spectrum (up to $\sim$300 TeV). The current design foresees 4 LSTs, 25 Davies Cotton (DC) MSTs, and 70 SSTs for CTA South which offers a privileged location to observe the Galactic Plane and Galactic Center over the full VHE range. The southern array may be even augmented with a proposed extension of up to 25 Schwarzschild-Couder (SC) MSTs. The northern site, with 4 LSTs and 15 MSTs, is expected to complement it. This very large number of telescopes improves the sensitivity (see Figure~\ref{fig:cta}) and the energy coverage by at least an order of magnitude compared to existing VHE instruments, as well as the angular resolution and the interval of energy detectable. The CTA collaboration now comprises more than 1200 members from 31 countries and the very large number of contributions at this year's ICRC is a direct sign that CTA is very well advanced. 

Prototypes are being assembled and the simulations and data analysis package is also moving forward. The science that will be available thanks to the excellent performances of CTA is huge: 
\begin{itemize}
\item with its large field of view, CTA will be an excellent experiment for surveys, 
\item with its fast slewing CTA will be ideal for transients, 
\item with its high energy coverage CTA will be excellent for PeVatron search, 
\item with its excellent angular resolution CTA will be excellent for morphological studies, 
\item with its excellent energy resolution CTA will be excellent for line search, 
\item with its energy coverage down to 20 GeV CTA will be excellent for cosmology. 
\end{itemize}

\subsection{The Large High Altitude Air Shower Observatory}
\begin{figure}[t]
 \centering
     \includegraphics[height=0.3\textheight]{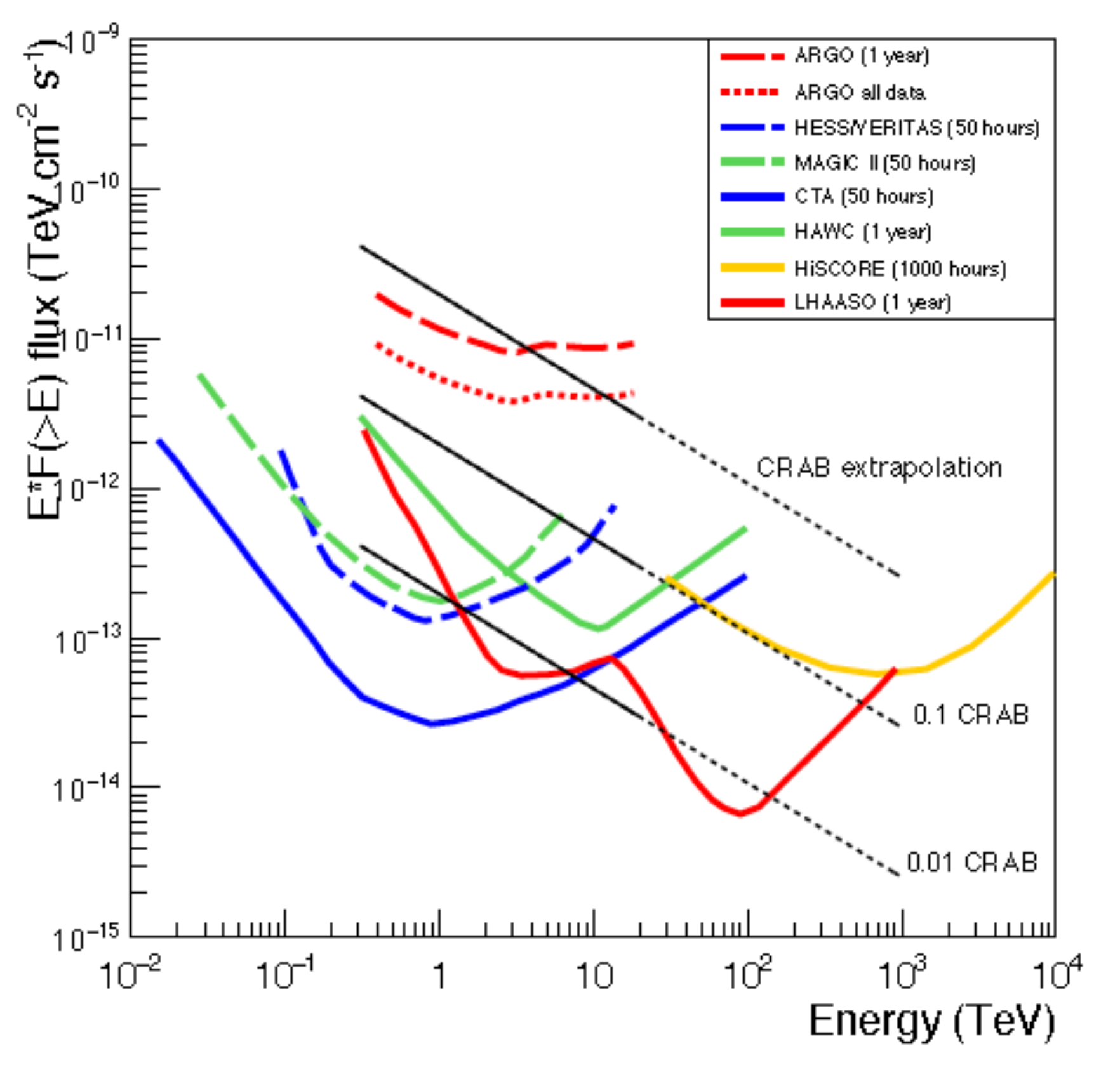}
     \includegraphics[height=0.3\textheight]{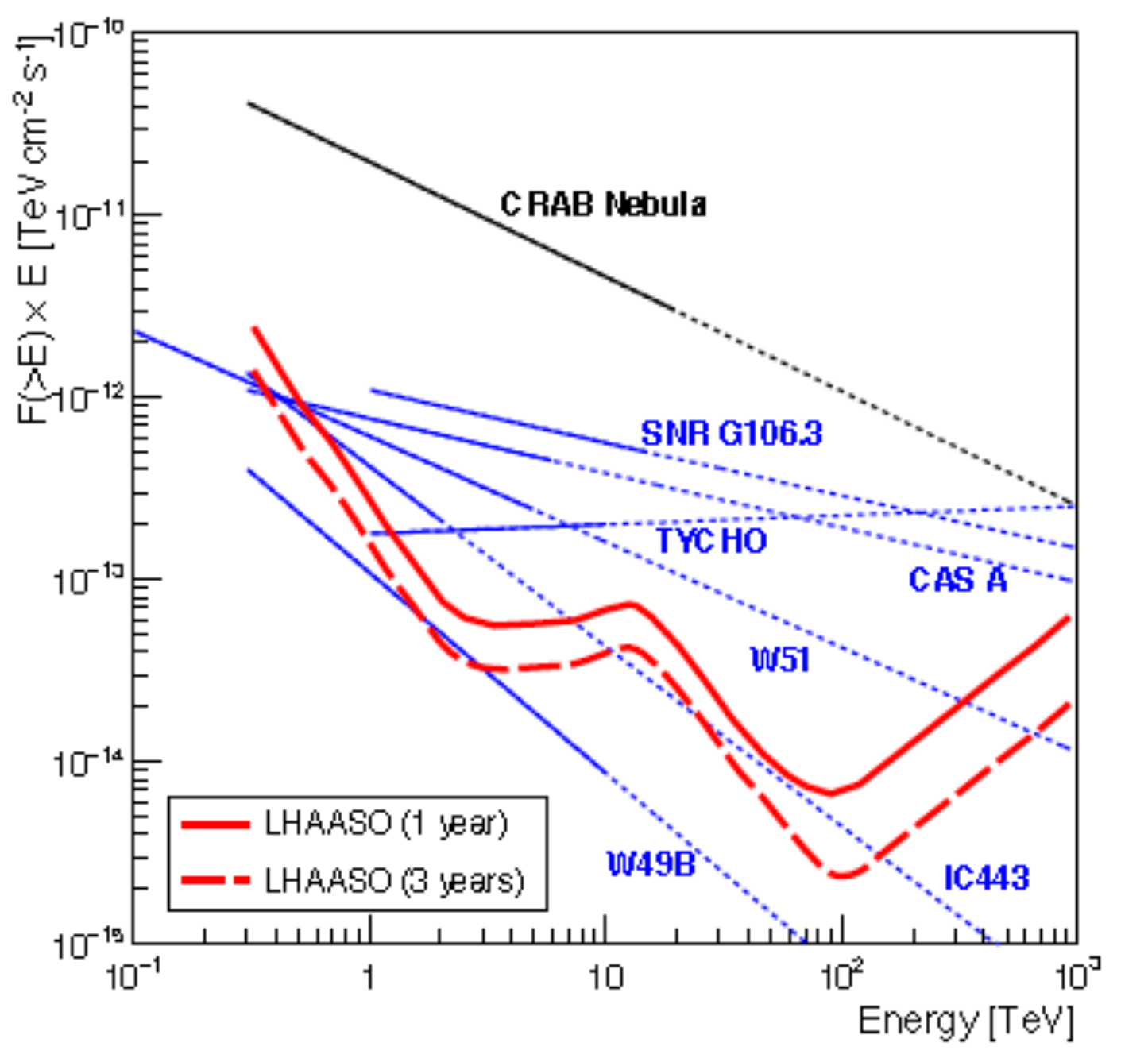}
  \caption{Left: Sensitivity of LHAASO to a Crab-like point gamma ray source compared to other experiments.
The Crab spectrum measured by ARGO-YBJ from 300 GeV to 20 TeV extrapolated to 1 PeV is used as a reference flux. Right: integral spectra of the six SNRs in the LHAASO field of view extrapolated to 1 PeV. The solid lines represent the measured spectra, while the extrapolations are shown by dotted lines.
}
  \label{fig:lhaaso}
\end{figure}

CTA is not the only way to go in the future. The Large High Altitude Air Shower Observatory (LHAASO) plans to build a hybrid extensive air shower (EAS) array at an altitude of 4410 m a.s.l. in Sichuan province, aiming for very high energy gamma ray astronomy and study of cosmic rays with energies in $10^{13}$ -- $10^{18}$ eV~\cite{lhaaso}. LHAASO consists of an extensive air shower array covering an area of 1 km$^2$ (KM2A), 75000 m$^{2}$ water Cherenkov detector array (WCDA), an in-filled shower core detector array (SCDA) and 12 wide-field air Cherenkov/fluorescence telescopes. An integral sensitivity of 1\% Crab unit can be reached at 3 TeV and 50 TeV as seen in Figure~\ref{fig:lhaaso} (Left) in comparison to the sensitivity curves of other detectors. Thanks to its large area and its high capability of background rejection, LHAASO can reach a sensitivity at energies above 30 TeV that is significantly higher than that of current instruments (and even 5 times higher than that of CTA), offering the possibility to monitor the gamma-ray sky up to PeV energies with an unprecedented sensitivity. In Figure~\ref{fig:lhaaso} (Right), the sensitivity of LHAASO is compared with the extrapolation of fluxes of six known SNRs. Four out of these six SNRs (Tycho, CasA, SNR G106.3+2.7 and W51) have fluxes higher than the LHAASO one-year sensitivity~\cite{lhaaso_galactic}.\\
It seems clear that projects using Cherenkov telescopes or using particle detector are very complementary since each of them are exploring different aspects of the gamma ray emission. Below 10 TeV, observing a single source, a Cherenkov telescope array as CTA has a higher sensitivity compared to particle detectors like HAWC and LHAASO. Thanks to the better angular and energy resolution, a Cherenkov telescope can perform detailed morphological and spectral analyses. Particle detectors, however, can monitor a source throughout the year and, thanks to their large field of view, have a much bigger chance to catch unpredictable transient events like flares. In addition, the energy coverage of LHAASO and CTA is obviously very different. Prototype detector arrays of about 1\% LHAASO are designed and built at the Yangbajing Cosmic Ray Observatory and used to coincidentally measure cosmic rays with the ARGO-YBJ experiment. The LHAASO field preparation will start in 2015 and detector construction is expected to start in 2016 and finish in 2020. According to the construction schedule, 1/4 of LHAASO will be put into operation and produce physical data in 2018.

Abother on-going project is called M@TE (Monitoring at TeV Energies). Its goal is to build a SiPM camera for one of the telescope and join the blazar monitoring to better catch the rising and falling edge of a flare and perform more continuous monitorings. In Mexico, there are two telescope mounts available. Other possible sites, to close the gaps even more, are India, Japan and Hawaii. Given the dropping prices of the photosensors, these extensions for the long-term monitoring have become affordable small-scale instruments~\cite{dorner}.

Finally, MACHETE~\cite{machete}, is a concept proposing to build an array of two non-steerable telescopes with a FOV of 5$\times$60 sq.deg. oriented along the meridian. With such configuration, roughly half of the sky drifts through the FOV in a year. The sensitivity that MACHETE would achieve after 5 years of operation for every source in this half of the sky is comparable to the sensitivity that a current IACT achieves for a specific source after a 50 h devoted observation. In addition, for sources observable in a single night, it reaches a sensitivity of 8\% Crab which is perfect to trigger other telescopes.

\section{Concluding remarks}
Every ICRC over the last decade has seen dramatic advances in ground-based gamma-ray astronomy, and this meeting
was no exception. All collaborations have produced a wealth of exciting results. One cannot summarize easily (and objectively) the 300 contributions that have been presented during this week since the number of discoveries that have been announced is quite impressive: first detection of a superbubble (in the LMC), first evidence of the detection of a PeVatron at the Galactic Center, first morphological study of a SNR interacting with a molecular cloud, first detection of pulsed emission from the Crab pulsar above 1 TeV, first gravitationally lensed blazar detected at the VHE energies, detection of the most distant blazar at TeV energies with a redshift of z=0.939... These beautiful results are an excellent demonstration that VHE phenomena are ubiquitous throughout the Universe. But many of the results have raised new questions which require more and better data for a deeper understanding of the underlying phenomena. Figure~\ref{fig:tevsrcs} very well shows that that the only limiting factor in our field is not the number of sources but only the sensitivity of the instruments. New projects are therefore crucial and are already very well advanced, in particular CTA and LHAASO. By the next ICRC, the deployment of the instruments for these 2 projects will have started, marking another important step in the development of the field. I look forward to seeing you all there.

\section*{Acknowledgments}
I would like to thank the organizers for the invitation to give a rapporteur talk at the 34th ICRC. I also want to thank all the collaborations for making the presentations available beforehand and for their support during this ICRC. Thank you to all the TeV colleagues for sharing and explaining all your great results before, during and after the talks. I really appreciated the discussion that we had during this week.

\end{document}